\documentclass[prl,twocolumn,amssymb,floatfix,superscriptaddress]{revtex4-2}
\usepackage{ulem}

\usepackage{graphicx}
\usepackage{amsmath}
\usepackage{amstext}
\usepackage{amssymb}
\usepackage[colorlinks,citecolor=blue]{hyperref}
\usepackage{graphicx}
\usepackage{tabularx}
\usepackage{array}
\usepackage{amsmath}
\usepackage{amstext}
\usepackage{longtable}
\usepackage{amssymb}
\usepackage{amsfonts}
\usepackage{longtable,booktabs}
\usepackage{hyperref}\usepackage{url}
\usepackage{dsfont,bbm}
\usepackage[usenames,dvipsnames]{color}
\usepackage{amsbsy}
\usepackage{dcolumn}
\usepackage{amsthm}
\usepackage{bm}
\usepackage{tikz}
\usepackage{multirow}
\usepackage{hyperref}
\hypersetup{
    colorlinks=true,
    linkcolor=black,
    filecolor=black,
    urlcolor=black,
}

\usepackage{mathrsfs}
\usepackage{amsfonts}
\usepackage{amsbsy}
\usepackage{dcolumn}
\usepackage{bm}
\usepackage{multirow}
\usepackage{color}
\usepackage{ulem}

\newcommand{\ua}{\uparrow} 
\newcommand{\da}{\downarrow}

\newcommand{\bk}{\mathbf k}

\newcommand{\bq}{\mathbf q}
\newcommand{\br}{\mathbf r}

\newcommand\mk{\boldsymbol{\mathrm{k}}}

\begin{document}

	\title{A 
    ``negative" route to pair density wave order}
    \author{Hao-Xin Wang}
    \thanks{These two authors contributed equally to this work.}
	\address{Department of Physics, The Chinese University of Hong Kong, Sha Tin, New Territories, Hong Kong, China}
	\author{Yi-Jian Hu} 
    \thanks{These two authors contributed equally to this work.}
    \address{Shenzhen Institute for Quantum Science and Engineering, Southern University of Science and Technology, Shenzhen 518055, Guangdong, China}
    \address{School of Physical Sciences, Great Bay University, Dongguan 523000, China}
    \address{International Quantum Academy, Shenzhen 518048, China}
	\author{Wen Huang}
	\email{huangwen@gbu.edu.cn}
        \address{School of Physical Sciences, Great Bay University, Dongguan 523000, China}
        \address{Shenzhen Institute for Quantum Science and Engineering, Southern University of Science and Technology, Shenzhen 518055, Guangdong, China}
        \address{International Quantum Academy, Shenzhen 518048, China}
        \author{Hong Yao}
        \email{yaohong@tsinghua.edu.cn}
        \address{Institute for Advanced Study, Tsinghua University, Beijing 100084, China}

\date{\today}
\begin{abstract}
Pair density waves (PDW) are novel forms of superconducting states that exhibit periodically modulated pairing. A remaining challenge 
is to elucidate how intrinsic PDW order can emerge robustly in strongly correlated electrons. 
Here we propose that PDW is prone to form in strongly coupled multiband superconductors simply with interband Cooper pairing between electrons from oppositely dispersing bands.
This scenario is heuristically motivated by the observation that uniform interband pairing in such systems would exhibit negative superfluid weight---a signature of an instability towards pairing modulation, implying that PDW emerges naturally in the true ground state. Using large-scale density-matrix-renormalization-group calculations with finite-size scaling analysis, we demonstrate this PDW mechanism in a minimal model with strong interband attractions. Our simulations reveal power-law superconducting correlations characterized by incommensurate modulations. 
The exponent $K_{sc}$ of the power-law PDW correlation decreases systematically with increasing ladder width, confirming a genuine long-range PDW order in the 2D limit. 
Our study therefore demonstrates a promising route to robust PDW states in multiband systems.
\end{abstract}

\maketitle

{\bf Introduction.-} The pair density wave (PDW) order is a novel superconducting phenomenon where the Cooper pairing order parameter exhibits periodic spatial modulation. The well-known Fulde-Ferrell and Larkin-Ovchinnikov states~\cite{fulde1964superconductivity,larkin1965nonuniform}, which may appear in conventional superconductors subject to Zeeman fields, are prototypical examples of PDW phases. In the recent decade, much attention has been drawn to intrinsic PDW orders formed in superconductors with strong electron correlations~\cite{Fradkin:15,Agterberg:20}. There, PDW can emerge without the Zeeman splitting as the driving force. Most notable is the underdoped cuprate superconductors, where multiple experiments have revealed the existence of PDW order~\cite{Hamidian:16,Ruan:18,Edkins:19,DuZ:20,LiX:21,lee2023pairdensity}. Much effort has been devoted to the potential implications of the PDW for the enigmatic pseudogap phase, and, ultimately, for the high-$T_c$ Cooper pairing~\cite{Fradkin:15,Agterberg:20}. More recently, PDW has also been reported in unconventional superconductors beyond cuprates, including Fe-based superconductors~\cite{liuPairDensityWave2023,zhang2024visualizing}, UTe$_2$~\cite{gu2023detection},  $A$V$_3$Sb$_5$~\cite{chen2021roton,deng2024chiral}, and transition-metal dichalcogenides~\cite{Xiaolong2021Discovery_Science, LiXuan2025Unidirectional}. On the theory front, it has proven extremely difficult to elucidate how PDW emerges in such strongly correlated systems amid potential close competition with other novel electronic orders \cite{Berg2010,Jaefari2012,SKJian-2015PRL,ZYHan-2020PRL,Shaffer2023Triplet,Shaffer2023Weak,Wu2023PairPRB,Castro2023Emergence,liu2024enhanced,ticea2024pair,Chen2023Singlet,Tsvelik2023,Wu2023Sublattice,Han2022Pair,yue2024pseudogap,Donna:23dwave,Patrick2014Amperean,Wu2023Pair,Haokai2023,Setty2023Exact,Jiang2023Pair,JiangAndTom2023Pair-Frontiers,Jin2022PRL,Banerjee2022Charge,Song2022Doping,GuoHM2024,Kivelson2020NPJQuantumM, Jiucai2025Pair, Yifan2024Pair, Yahui2022Pair}. 

\begin{figure}[h]
\includegraphics[width=8.5cm]{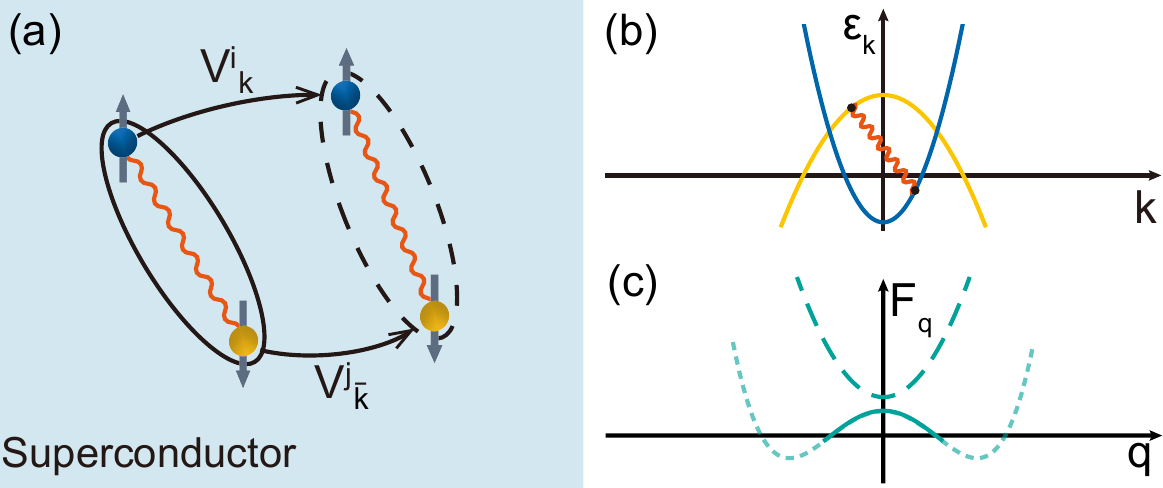}
\caption{(Color online) (a) Illustration of an effective self-Josephson effect in a superconductor, generated by transfer of Cooper pairs. Superfluid weight is a measure of this Josephson coupling. Cooper pairing is depicted by wiggly lines encircled by an ellipse, with one electron (blue sphere) from band-$i$ with wavevector $\mk$ and another (orange sphere) from band-$j$ with $\bar{\mk}=-\mk$. The transfer of the Cooper pair is facilitated by the motion of the two electrons, described respectively by their velocities $V^i_{\mk}$ and $V^j_{\bar{\mk}}$. (b) Interband Cooper pairing between two oppositely dispersing bands. (c) Free energy as a function of the superconducting phase modulation wavevector $\bq$. The two curves respectively describe the two scenarios where uniform pairing (i.e. $\bq=0$) would exhibit positive (upper curve) and negative (lower curve) superfluid weight.}
\label{fig:illustration}
\end{figure}

In conventional wisdom, spatial modulations of the superconducting order parameter tend to cost energy. For example, a phase modulation of $\Delta e^{i\theta}$ would incur an energy cost of 
\begin{equation}
    F_\theta = D |\nabla \theta|^2 + \mathcal{O}(|\nabla \theta|^2)\,,
\end{equation}
where $D$ denotes the superfluid weight (SW) and the second term represents higher order contributions. In typical scenarios, $D>0$, meaning that the pairing is stable against phase modulations. Recently, it was proposed that $D<0$ may be realized in some multiband superconductors by virtue of the so-called quantum geometric effects~\cite{chenPairDensityWave2023,wang2024quantum,jiangPairDensityWaves2023,kitamuraQuantumGeometricEffect2022}. Already at this level of consideration, this sets a favorable condition for the formation of a superconducting order exhibiting phase modulations~\cite{chenPairDensityWave2023,wang2024quantum,jiangPairDensityWaves2023,kitamuraQuantumGeometricEffect2022,ticea2024pair,sun2024flatbandfuldeferrelllarkinovchinnikovstatequantum,han2024quantum}, as illustrated in Fig.~\ref{fig:illustration} (c). 

Heuristically, the SW measures the strength of an effective self-Josephson coupling in a superconductor. In analogy to the standard Josephson effect which is induced by tunneling Cooper pairs between two superconductors placed in adjacency, this self-Josephson coupling is generated by the transfer of Cooper pairs across a uniform superconductor, as schematically illustrated in Fig.~\ref{fig:illustration}. Such processes are enabled by the motion of electrons, which is described by their velocity operators $V_{\mu\mk}^{i}c^\dagger_{i\mk \ua}c_{i\mk \ua}$ and $V_{\mu\bar{\mk}}^{j}c^\dagger_{j\bar{\mk} \da}c_{j\bar{\mk}\da}$. Here, $i$ stands for the band index, and $\mu$ labels the component of the velocity, and the velocities are simply the group velocities, i.e. $V_{\mu\mk }^{i}= \partial_{k_\mu}\epsilon_{i\mk}$. Indeed, a formal linear response derivation of the SW of a one-band $s$-wave superconductor would lead to a simple expression relating to the two velocities as follows~\cite{liangBandGeometryBerry2017, Hu2025Quantumgeometric} $D^i_{\mu\nu} = -\frac{1}{\Omega}\sum_{\mk}\frac{V_{\mu\mk}^{i} (V_{\nu\bar{\mk}}^{i})^\ast\Delta_{i}^\ast\Delta_{i} }{(E_{i\mk})^3}$, where $E_{i\mk}=\sqrt{\epsilon_{i\mk}^2+|\Delta_i|^2}$ and $\Omega$ denotes the volume of the system. Note that we have used $\Delta_i^\ast\Delta_i$, instead of $|\Delta_i|^2$, to manifest the Josephson effect analogy. Importantly, since $V_{\nu\bar{\mk}}^i=-V_{\nu\mk}^i$, this SW is nonnegative and, in the weak-coupling and continuum limit, its diagonal elements reproduce the well-known result of the carrier density divided by the effective mass~\cite{annett2004superconductivity,altland2010condensed}.

The same analogy also applies to models where Cooper pairing takes place between different bands, i.e. $\Delta_{ij} c_{i\bar{\mk}\ua}c_{j\mk\da}$ (see sketch in Fig.~\ref{fig:illustration}). Here, the SW generated by the transfer of these interband Cooper pairs are related to the $\mk$-integrals involving the following terms in the numerator of the integrand~\cite{Hu2025Quantumgeometric},
\begin{equation}
V^{i}_{\mu\mk}V^{j}_{\nu\mk}\Delta_{ij}^\ast \Delta_{ij}\,\,\,\, \text{and} \,\,\,
    \ V^{j}_{\mu\mk}V^{i}_{\nu\mk}\Delta_{ji}^\ast \Delta_{ji}\,.
    \label{eq:processd}
\end{equation}
Intriguingly, this SW (with $\mu=\nu=x$) is not necessarily positive-definite. Its sign is determined by the relative sign of the velocities of the two paired electrons. For a two-band model with oppositely dispersing bands, this SW becomes negative~\cite{Hu2025Quantumgeometric}! Whether this exotic scenario would lead to PDW order is the central theme of the present study. In this paper, we shall demonstrate that such a model with strong interband pairing interaction indeed tends to develop PDW order.

\begin{figure}
    \centering
    \includegraphics[width=\linewidth]{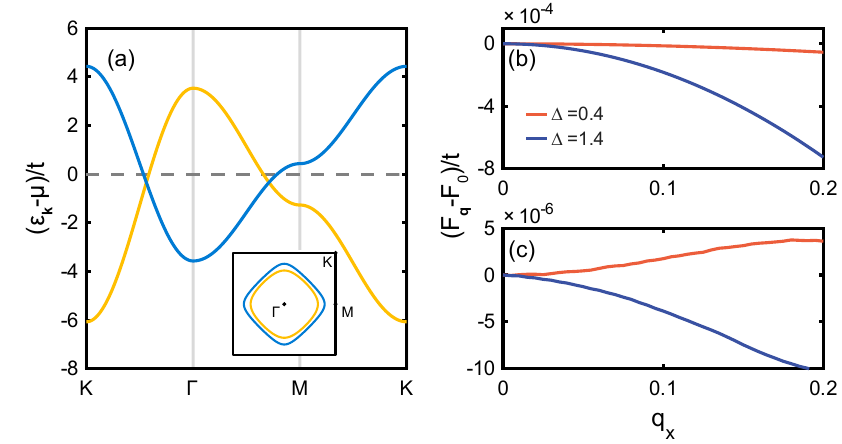}
    \caption{(a) The band structure of the bilayer model Eq.~\eqref{eq:hamk} with $(t_1,t_2)=(-1,1.2)t$, and the chemical potential $(\mu_1,\mu_2)$ are chosen so that the electron filling on the two bands are $x_1=0.4$ and $x_2=0.7$, respectively. The inset shows the Fermi surfaces. The right panel shows the variation of the free energy with modulation wavevector $\bq=(q_x,0)$ for (b) $s$-wave and (c) $d$-wave states, and with the pairing amplitude held fixed.}
    \label{fig:bandFq}
\end{figure}

{\bf Model and Method.-} We construct a spinless two-band model with kinetic Hamiltonian given by,
\begin{equation}
\mathcal{H}_{\mk}=\sum_{a,\mk}\left(\varepsilon_{a,\mk}-\mu_a\right)c_{a,\mk}^{\dagger}c_{a,\mk},
\label{eq:hamk}
\end{equation}
where $a=1,2$ label the bands, and $\varepsilon_{a,\mk}$ and $\mu_a$ stand respectively for the dispersion and chemical potential of the two bands. Absent interband mixing, no non-Abelian Berry phase exists connecting the wavefunction of the two bands, hence the model is quantum-geometry-free. We consider a simplified scenario where the system resides on a square lattice with only nearest-neighbor hopping, i.e.~$\varepsilon_{a,\mk}=2t_a(\cos{k_x}+\cos{k_y})$. Our conclusion nonetheless applies to models with further neighbor hoppings, as long as the opposite-dispersion condition is met. Here, simply setting $t_1t_2<0$ returns the desired two oppositely dispersing bands, irrespective of the filling on the two bands. Throughout this study, we choose appropriate chemical potentials so as to obtain a pair of electron-like and hole-like Fermi surfaces in the normal state, as shown in Fig.~\ref{fig:bandFq} (a). 



Consider now interband Cooper pairing with finite momentum $\bq$, $\Delta_{12,\bq} f_{\bk}c_{1,\bar{\bk}}c_{2,\bk+\bq}$, where $f_{\mk}$ denotes the form of the pairing function, e.g. $f_{\mk}\equiv 1$ for $s$-wave pairing and $f_{\mk} = \cos k_x - \cos k_y$ for $d$-wave. Note that, due to the Pauli principle, such interband pairings switch sign upon the exchange of band indices. For a preliminary test of the above heuristic argument, we assume that the pairing amplitude is independent of $\bq$, $\Delta_{12,\bq}\equiv \Delta_0$, and evaluate the mean-field zero-temperature free energy as a function of $\bq$~(see Supplementary Material (SM)~\cite{supp} for details). The results for $s$- and $d$-wave states for a representative set of parameters are presented in Fig.~\ref{fig:bandFq} (b) and (c). The free energy $F_{\bq}$ is expanded around $\bq=0$ as $F_{\bq} = D \bq^2 + \mathcal{O}(\bq^4)$. For the $s$-wave pairing, we find that $F_{\bq}$ is always a local maximum at $\bq=0$. That is, we obtain $D<0$ regardless of the magnitude of $\Delta_0$, thus confirming our expectation. 

The $d$-wave state behaves somewhat differently (Fig.~\ref{fig:bandFq} (c)). Its free energy exhibits a local minimum at $\bq=0$, i.e.~$D>0$, for relatively weaker interband pairing, yielding $D<0$ only above a critical pairing strength. The deviation at weak pairing is related to an additional contribution to the SW not described by the Cooper pair transfer processes in \eqref{eq:processd}. This contribution is associated with the energy cost when the Cooper pair wavefunction is deformed by a real-space phase gradient at finite $\bq$ (i.e.~$\Delta_{12}e^{i\theta(\br)}=\Delta_{12}e^{i\bq\cdot \br}$)~\cite{Hu2025Quantumgeometric}. It is absent for the simple $s$-wave state, but exists when the pairing takes place between electrons located on different sites, that is, when the form factor $f_{\bk}$ has explicit $\bk$-dependence. In this scenario, the two electrons of a Cooper pair may experience different phases when a phase gradient is applied. Nevertheless, the sign and magnitude of this contribution are sensitive to microscopic details of the form factor and those of the band structure. Under the set of parameters used in Fig.~\ref{fig:bandFq}, the Cooper pair transfer contribution dominates -- and the total SW turns negative -- at stronger pairing.  

The above simplified mean-field results are informative, but must be treated with caution. Due to the absence of inherent interband Cooper pairing instability, the interaction must exceed a critical strength typically of the order of bandwidth or even larger, making the model a strong coupling problem. To identify the true ground state, it is thus highly desirable to go beyond mean-field. In the following, we provide large-scale density matrix renormalization group (DMRG) calculations of a simple representation of the interband $s$-wave model in real space~\cite{White1992PRL,White1993PRB}. We adopt the following onsite interband density-density attraction
\begin{equation}
    \mathcal{H}_{\mathrm{int}}=-U\sum_{i}n_{i,1}n_{i,2}\,,
    \label{eq:HintDMRG}
\end{equation}
where $i$ labels the lattice sites. We consider a set of filling, $x_1 = 0.4, $ and $x_2=0.7$ on the two bands, with the corresponding Fermi surfaces in the noninteracting limit shown in Fig.~\ref{fig:bandFq} (a). This choice ensures that the results obtained are generic, free of complications arising, in particular, from Fermi surface nesting. To establish the generality of our findings, we have explored an extended parameter space, including $t_2$ values from 1 to 1.2 and interaction strengths $U$ from 2 to 8 (in units of $|t_1|=1$). Across all these parameter sets, the raw data consistently reveal power-law decay of superconducting correlations and divergent superconducting susceptibility at width $L_y=4$, suggesting the robustness of the phenomenon. For a subset of parameters, we have performed a meticulous finite-size scaling analysis to establish the results quantitatively in the 2D limit. In the following, we present in detail the results for $U=4$ and $t_2 = 1.2$, which serve as a representative case. Additional data for $U=4$, $t_2=1$ which underwent equally rigorous analysis are provided in the SM~\cite{supp}.

In the DMRG calculations, we use $L_x\times L_y$ lattices with open boundary in $x$-direction and periodic boundary along $y$. Setting $L_x \gg L_y$, we systematically approach the 2D limit by going from $L_y = 2,$ to 4. We kept the bond dimension up to $D=26000$ to ensure sufficient convergence, with largest truncation error in the order of $10^{-6}$.

{\bf PDW state from the two-band model--} The superconductivity is characterized by the pair-pair correlation function
\begin{equation}
    \Phi(\mathbf{r}) = \langle \Delta^{s\dagger}_i \Delta^s_j \rangle,
\end{equation} 
where the order parameter $\Delta^s_i = c_{2,i}c_{1,i}$ denotes the on-site interband $s$-wave pairing field, and $\mathbf{r} = (r_x, r_y)$ is the vector connecting the $i$ and $j$ lattice sites.

\begin{figure}[t]
\centering
\includegraphics[width=\linewidth]{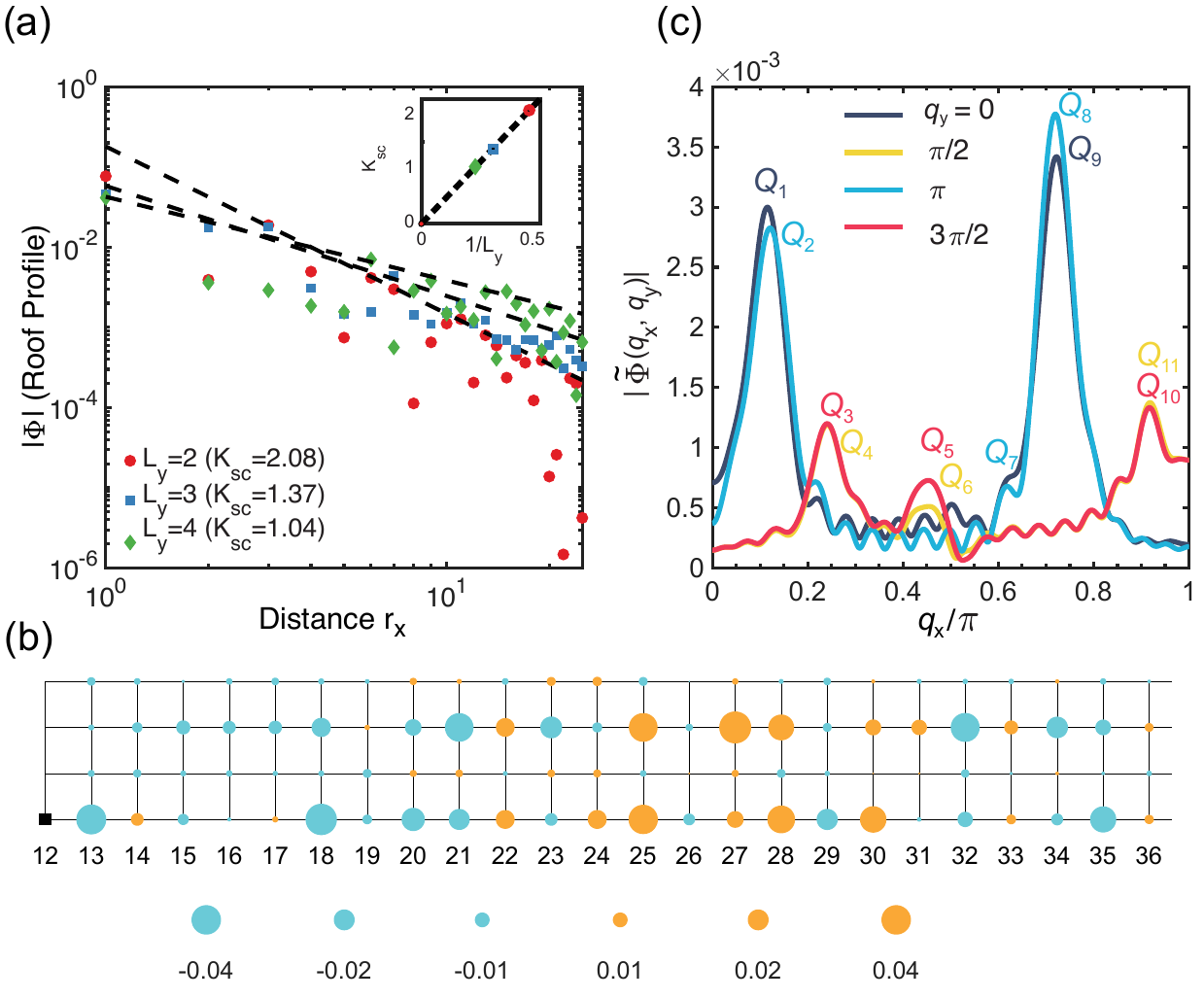}
\caption{
(a) Envelope of superconducting correlations $|\Phi(\mathbf{r})|$ as a function of distance $r_x$ for width $L_y=2,3,4$, with power-law fits $|\Phi(\mathbf{r})| \sim r_x^{-K_{sc}}$. 
(b) Normalized superconducting correlations $f(\mathbf{r}) = \Phi(\mathbf{r}) \cdot r_x^{K_{sc}}$ for $L_y=4$, revealing the spatial modulation characteristic of a pair-density-wave. Positive (negative) values are represented by orange (blue) circles, with areas proportional to their magnitudes. The reference point is marked by a black square. The exponent $K_{sc}$ used for normalization is determined from the power-law fits in (a). 
(c) Fourier transform of the normalized superconducting correlation data. Only half of the Brillouin zone is shown, because $|\Tilde{\Phi}(\mathbf{q})| = |\Tilde{\Phi}(-\mathbf{q})|$.}
\label{fig:DMRG_SC_corr2}
\end{figure}

Figure~\ref{fig:DMRG_SC_corr2} (a) shows the superconducting correlations obtained in DMRG. The horizontal axis represents the $x$-component of $\br$. To mitigate the complexity arising from modulations along the $y$-direction and to clearly reveal the decay profile, we plot the maximum absolute value of $\Phi(\mathbf{r})$ for each $r_x$. For all $L_y$'s, the correlation exhibits a power-law decay $|\Phi(\mathbf{r})| \propto r_x^{-K_{sc}}$. The fitted exponent $K_{sc}$ decreases from 2.08 at $L_y=2$ to 1.04 at $L_y=4$. Moreover, the exponent scales as $K_{sc} \propto 1/L_y$, as shown in the inset of Figure~\ref{fig:DMRG_SC_corr2} (a). This strongly suggests the development of off-diagonal long-range order (characteristic of superconductivity) in the 2D limit \cite{Yang1962Concept}.


Having established the superconducting order, we now examine its spatial modulation. Rather than directly showing the real-space correlation function $\Phi(\mathbf{r})$, we present the normalized correlations $f(\mathbf{r}) = \Phi(\mathbf{r}) \cdot r_x^{K_{sc}}$, with the background power-law decay removed. Figure \ref{fig:DMRG_SC_corr2} (b) displays $f(\mathbf{r})$ for $L_y=4$. Despite employing complex-valued DMRG, all computed correlations are real-valued. The pattern clearly reveals a pairing field whose sign and amplitude both modulate in real space. Although data for $L_y=2$ and $3$ are not shown here, they all exhibit qualitatively similar modulation pattern.

To quantify the PDW momenta, we perform a Fourier analysis on the normalized correlations for the $L_y=4$ system [Fig.~\ref{fig:DMRG_SC_corr2} (d)], using the definition:
\begin{equation}
    \tilde{\Phi}(\mathbf{q}) = \frac{1}{N}\sum_{\mathbf{r}_j} f(\mathbf{r}_0 - \mathbf{r}_j) e^{-i (\mathbf{r}_0 - \mathbf{r}_j) \cdot \mathbf{q}},
\end{equation}
where $\mathbf{r}_0$ is a fixed reference site. Note that $k_y$ is limited to discrete values $n\pi/L_y$ ($n=0, \ldots, L_y-1$). Only half of the Brillouin zone ($k_x \in [0, \pi]$) is shown; the other half is related by mirror symmetry due to $|\Tilde{\Phi}(\mathbf{q})| = |\Tilde{\Phi}(-\mathbf{q})|$, which implies the degeneracy of PDWs with wavevectors $\pm \mathbf{Q}$. The Fourier spectrum exhibits prominent peaks at several wavevectors, while also clearly showing weak intensity at zero momentum. This indicates that the SC in the ground state does not contain significant uniform component and is thus primarily finite-momentum pairing. 

\begin{figure}[tbp]
    \centering
    \includegraphics[width=\linewidth]{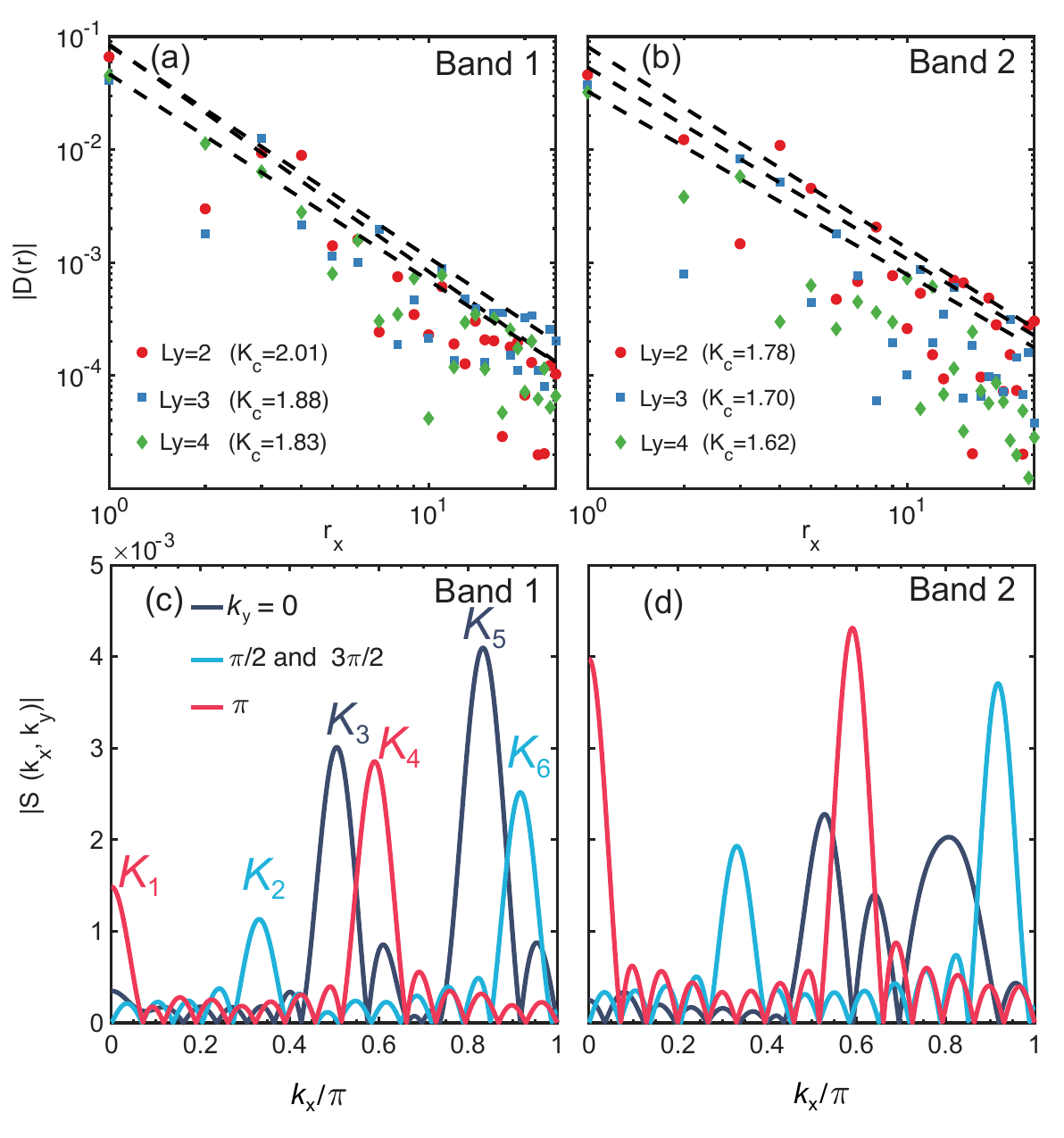}
    \caption{(a-b) Intraband density correlation $|D(r)|$ for $L_y = 2,3,4$ with power-law fits $|D(r)| \sim r_x^{-K_c}$. The dashed lines are fitting lines. (c-d)
    Density structure factor $|S(\mathbf{k})|$ in the reduced Brillouin zone 
             ($k_x \in [0,\pi]$) exploiting $|S(\mathbf{k})| = |S(\mathbf{-k})|$.  Data at $k_y = \pi/2$ and $k_y = 3\pi/2$ (blue curves) are scaled 3$\times$ for visual clarity. To minimize boundary effects, we exclude edge data in the Fourier analysis. Primary CDW peaks are labeled $K_1$ to $K_{6}$.}
    \label{fig: Density structure factor}
\end{figure}

To check that PDW is the primary order, we also investigate the CDW order by calculating the charge density correlations
\begin{equation}
    D_{a,b}(\mathbf{r})= \langle n_{a,i} n_{b,j} \rangle,
\end{equation}
and analyzing the charge density modulations. The profiles of the intraband charge density correlations $D_{11}(\mathbf{r})$ and $D_{22}(\mathbf{r})$ in Fig.~\ref{fig: Density structure factor} (a) invariably display power-law decay. The associated exponents $K_c$ are obtained from fits to $|D_{a,b}(\mathbf{r})| \propto r_x^{-K_c}$. However, unlike the systematic decrease of the PDW exponent $K_{sc}$ with $L_y$ shown in Fig.~\ref{fig:DMRG_SC_corr2} (b), the CDW exponent $K_c$ does not decrease as significantly. This indicates that the PDW instability becomes increasingly dominant over the CDW as the system width grows. This observation establishes PDW as an intrinsic instablity in the thermodynamic limit, and rules out scenarios where PDW emerges as a secondary effect, e.g., from uniform superconductivity coupled to CDW. 

The simultaneous enhancement of PDW and CDW correlations with increasing $L_y$ differs from the behavior in Luther-Emery liquids, where CDW and SC instabilities evolve in opposite trends as the system width increases \cite{Giamarchi_book, Gannot2023How}. This distinction can be explained if the CDW in our model arises in association with composites of the PDWs. A direct evidence is that the CDW wavevectors are determined by those of the PDW. Specifically, the composite of two PDWs with wavevectors $Q_i$ and $Q_j$ can induce a CDW with modulation wavevector $K = Q_j - Q_i$, i.e., $\rho_K \propto \Delta_{Q_j}^\ast\Delta_{Q_i}$. To determine the CDW wavevectors in our calculation, we analyze the density structure factor $S_a(\mathbf{k})$ [Fig.~\ref{fig: Density structure factor} (c–d)], defined as:
\begin{equation}
S_{a}(\mathbf{k}) = \frac{1}{N} \sum_i (n_{a, i} - n_{a,0}) e^{-i \mathbf{k} \cdot \mathbf{r}_i},
\end{equation}
where $a$ labels the band and $n_{a,0}$ is the uniform background density. We exclude boundary data in the Fourier transform to minimize finite-size effects. After filtering out residual minor peaks, we identify several most prominent CDW wavevectors $\mathbf{K}_i$ within half of the Brillouin zone, leveraging the symmetry $S_a(\mathbf{k}) = S_a^*(\mathbf{-k})$. These CDW wavevectors $\mathbf{K}_i$, together with the PDW wavevectors $\mathbf{Q}_i$ identified in Fig.~\ref{fig:DMRG_SC_corr2} (c) (where, by symmetry, wavevectors $-\mathbf{K}_i$ and $-\mathbf{Q}_i$ are implicitly included), allow for a direct comparison. Remarkably, all major CDW peaks $\mathbf{K}_i$ can be traced to the PDW wavevectors by the relation $\mathbf{K}_i = \mathbf{Q}_j - \mathbf{Q}_k$, within a numerical tolerance of $0.02\pi$ set by finite-size effects. Representative sets of $(\mathbf{K}_i, \mathbf{Q}_j-\mathbf{Q}_k)$ satisfying this relation are tabulated in Table 1. This exact correspondence, combined with the dominant PDW correlations, provides compelling evidence that the CDW is not an independent competing instability, but a secondary order subordinate to the primary PDW order.

\begin{table}[t]
  \caption{Representative PDW wavevector pairs generating the six CDW wavevectors (in units of $\pi$).}
  \centering
  \begin{tabular}{c c c}
    \hline\hline
    CDW momenta & $\mathbf{K}=(K_x,K_y)$ & Representative pair $(\mathbf{Q}_\alpha,\mathbf{Q}_\beta)$ \\
    \hline
    $\mathbf{K}_1$ & $(0,\,1)$   & $\mathbf{Q}_4 - \mathbf{Q}_3$ \\
    $\mathbf{K}_2$ & $(0.331,\,0.5)$   & $\mathbf{Q}_6 - \mathbf{Q}_2$ \\
    $\mathbf{K}_3$ & $(0.517,\,0)$   & $\mathbf{Q}_7 - \mathbf{Q}_2$ \\
    $\mathbf{K}_4$ & $(0.591,\,1)$   & $\mathbf{Q}_8 - \mathbf{Q}_2$ \\
    $\mathbf{K}_5$ & $(0.821,\,0)$   & $\mathbf{Q}_8 - (-\mathbf{Q}_1)$ \\
    $\mathbf{K}_6$ & $(0.918,\,0.5)$   & $\mathbf{Q}_7 - (-\mathbf{Q}_5)$ \\
    \hline\hline
  \end{tabular}
\end{table}

{\bf Concluding remarks.-} 
We have argued that uniform interband Cooper pairing between two oppositely dispersing bands is inherently unstable towards PDW order due to the negative superfluid weight of the uniform pairing. Through large-scale DMRG calculations on such a two-band model with strong interband attraction, we have demonstrated that in the two-dimensional limit, the system develops off-diagonal long-range order with multiple finite modulation wavevectors. Our results unambiguously confirm a PDW ground state, characterized by incommensurate wavevectors. This primary PDW order, in turn, induces a subleading CDW order, as evidenced by the precise correspondence between their modulation wavevectors. 


We note that our PDW differs from the conventional field-induced FFLO states, which are weak-coupling phenomena and which, owing to the relatively small Zeeman splitting of the Fermi surface, typically exhibit small modulation wavevectors. In some special limits, our model reduces to models 
studied previously in two separate contexts. Ref. \onlinecite{Paramekanti2010prb} focused on insulating two-band models and found PDW with commensurate wavevector ${\mathbf Q}=(\pi,\pi)$ on a square lattice; Ref. \onlinecite{GuoHM2025prb} studied modified versions of the Hubbard model where the two spins acquire opposite hopping integrals, and found near half-filling $(\pi,\pi)$-PDW closely related to a peculiar Fermi surface nesting feature in the pairing channel. Our current study represents a significant advance over these ingenious works. In particular, we have provided a new intuitive insight into the driving mechanism of PDW and have demonstrated its validity for more generic two-band models with oppositely dispersing bands. 

It remains to examine whether the proposed mechanism is related to the PDW experimentally observed in a number of unconventional superconductors~\cite{Hamidian:16,Ruan:18,Edkins:19,DuZ:20,LiX:21,lee2023pairdensity,liuPairDensityWave2023,zhang2024visualizing,gu2023detection,chen2021roton,deng2024chiral}. Arguably, all existing superconductors are multiband in nature. Even when only one Bloch band crosses the Fermi level, it is usually derived from multiple hybridizing orbitals. However, in real materials the intraband Cooper pairing, which typically favors uniform pairing, dominates over interband pairing, thereby hindering the formation of PDW. On the other hand, our model may be simulated by cold atom systems where the band structure and interactions can be manipulated with relative ease. We expect our proposal to stimulate many further investigations on both superconductivity and cold-atom fronts. 







{\bf Acknowledgments} We acknowledge the helpful discussions with Shou-Shu Gong and Wen Sun. This work is supported by NSFC under Grants No.~12374042 and No.~11904155, the Guangdong Science and Technology Department under Grant 2022A1515011948, the Dongguan Key Laboratory of Artificial Intelligence Design for Advanced Materials, and a Shenzhen Science and Technology Program (Grant No.~KQTD20200820113010023). Computing resources are provided by the SongShan Lake HPC Center (SSL-HPC) in Great Bay University and the Center for Computational Science and Engineering at Southern University of Science and Technology.

\bibliography{huyijian}

\end{document}


\title{Supplementary Materials:\\
A ``negative" route to pair density wave order} 	
 
\maketitle

\section{Mean-Field theory}

The BdG Hamiltonian of our spinless two-band model with finite momentum pairing can be written as
\begin{equation}
    \mathcal{H}_{\mathrm{BdG}}=\Psi^\dagger_{\mk}
    \begin{pmatrix}
    \mathcal{H}_{\mk+\bq} & \tilde{\bDelta}_{\bq} \\
    \tilde{\bDelta}_{\bq} & -\mathcal{H}_{\bar{\mk}}
    \end{pmatrix} \Psi_{\mk},
\end{equation}
where $\Psi_{\mk}=(c_{1\mk+\bq},c_{2\mk+\bq},c^\dagger_{1\bar{\mk}},c^\dagger_{2\bar{\mk}})^T$, $\bar{\mk}=-\mk$, and $\mathcal{H}_{\mk}$ is the kinetic Hamiltonian, which is diagonal due to the absence of hybrization between the two bands. We consider a scenario where only interband pairing with finite momentum $\bq$ is formed, $\Delta_{12,\bq} f_{\bk}c_{1,\bar{\bk}}c_{2,\bk+\bq}$. The form factor $f_{\mk}=1,\cos{k_x}-\cos{k_y}$ describe interband $s$-wave and $d$-wave pairings, respectively. Hence, the pairing matrix reads:
\begin{equation}
    \tilde{\bDelta}=
    \begin{pmatrix}
        0 & \Delta_{12,\bq}\\
        \Delta_{12,\bq} & 0
    \end{pmatrix}.
\end{equation}
In the calculation of maintext, we assume the pairing amplitude is constant for small $\bq$, $\Delta_{12,\bq}\equiv \Delta_0$. The free energy is given by
\begin{equation}
\begin{aligned}
    F_\bq&=\frac{1}{\Omega}\sum_{a,\mk}\sum_{E_{a}<0}E_a(\mk,\bq)+\frac{1}{\Omega}\sum_{a,\mk}(\varepsilon_{a,\mk}-\mu_a)-\frac{1}{\Omega}\sum_{\mk}\frac{\Delta_0^2}{V}f_{\mk}^2,
    \end{aligned}
\end{equation}
where $E_a(\mk,\bq)$ is the quasi-particle dispersion of BdG Hamiltonian, $a=1,2$ is the band index, and $\Omega$ is the size of $k$-space. The last two terms are independent of $\bq$, so they will be canceled when we evaluate the free energy difference $F_{\bq}-F_{0}$, making the final result independent of the interaction strength $V$.\par

\section{Additional DMRG results}

This section presents additional DMRG results for parameters $-t_1 = t_2 = 1$, with all other settings (e.g., band fillings) consistent with those in the main text. These data, which include superconducting and charge density correlations, serve to demonstrate the robustness of our findings. The observation of a dominant PDW instability across different parameters confirms that the PDW state is not a fine-tuned artifact but a widespread phenomenon within our proposed negative-route mechanism.

\begin{figure}[t]
\centering
\includegraphics[width=\linewidth]{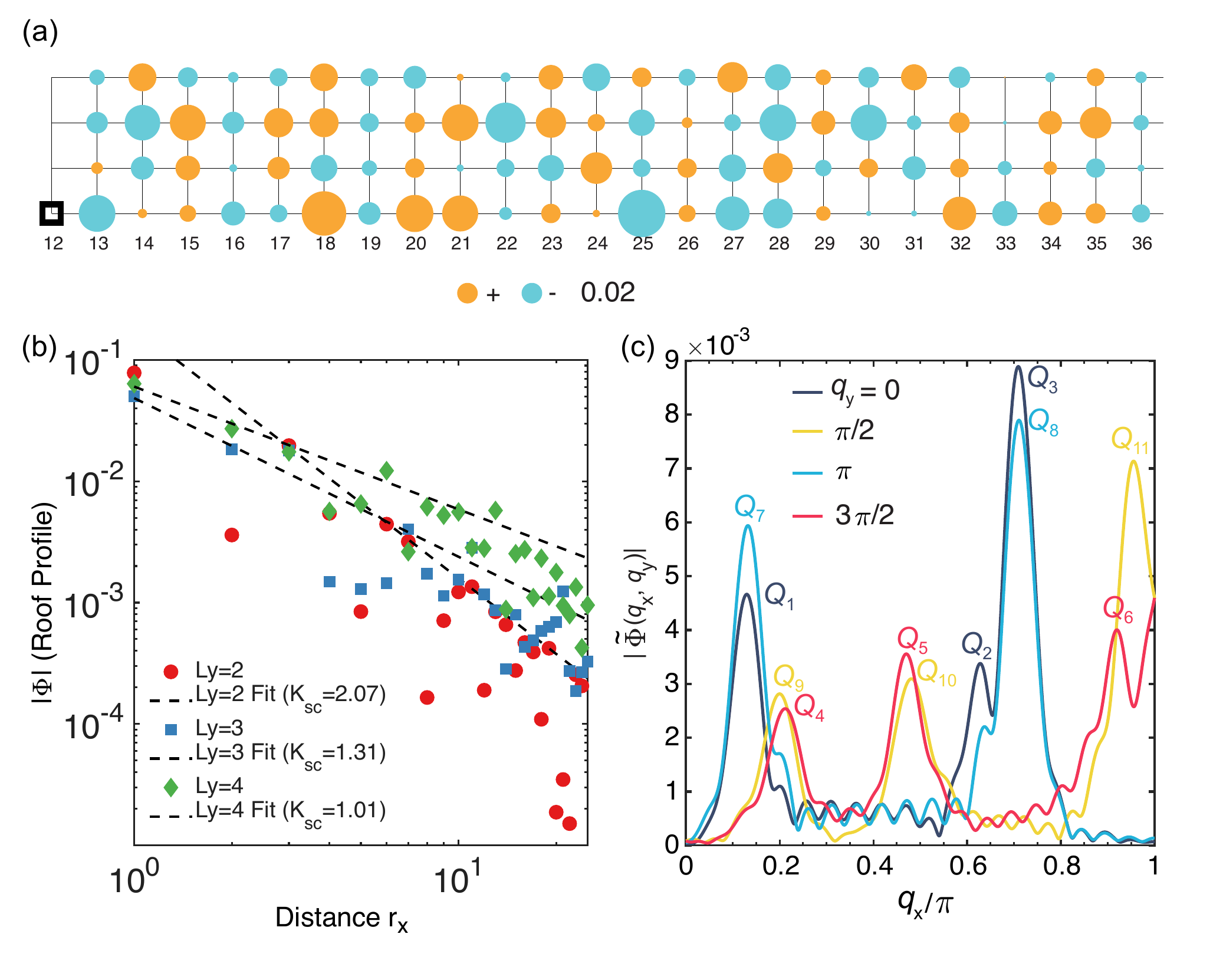}
\caption{(a) Normalized superconducting correlations $\Phi(\mathbf{r}) \cdot r_x^{K_{sc}}$ for $L_y=4$ showing the spatial modulation characteristic of a pair-density-wave. Positive (negative) values are shown as orange (blue) circles, with area proportional to magnitude. The reference point is marked as the black square. The envelope function $r_x^{K_{sc}}$ is obtained in (b). 
(b) Envelope of superconducting correlations $|\Phi(\mathbf{r})|$ versus distance for $L_y=2,3,4$ with power-law fits $|\Phi(\mathbf{r})| \sim r_x^{-K_{sc}}$. (c) Fourier transformation of the normalized superconducting correlation data. Only half of the Brillion zone is presented because $|\tilde{\Phi}(\mathbf{q})| = |\tilde{\Phi}(-\mathbf{q})|$.}
\label{Sfig:DMRG_SC_corr2}
\end{figure}

Figure \ref{Sfig:DMRG_SC_corr2}(b) shows the envelope of the superconducting correlations $|\Phi(\mathbf{r})|$ along the $x$-direction, obtained by taking the maximum absolute value at each $r_x$ to mitigate $y$-direction modulations. For all system widths, the correlations exhibit a clear power-law decay $|\Phi(\mathbf{r})| \sim r_x^{-K_{sc}}$. The fitted exponents $K_{sc}$ decrease from $2.07$ for $L_y=2$ to $1.01$ for $L_y=4$. Values of $K_{sc} < 2$ for the 3- and 4-leg cylinders imply a divergent pairing susceptibility in the thermodynamic limit. Furthermore, these exponents closely follow the scaling relation $K_{sc} \propto 1/L_y$ , reinforcing the emergence of off-diagonal long-range order in the two-dimensional (2D) limit.

The real-space modulation of the superconducting order is visualized in Figure \ref{Sfig:DMRG_SC_corr2}(a), which plots the normalized correlations $f(\mathbf{r}) = \Phi(\mathbf{r}) \cdot r_x^{K_{sc}}$ to remove the power-law decay background. The observed periodic sign reversals are a hallmark of PDW order. This is quantified by the Fourier analysis in Figure \ref{Sfig:DMRG_SC_corr2}(c), which reveals prominent peaks at finite momenta and a suppressed spectral weight at zero momentum, unequivocally identifying the superconductivity as a PDW. Multiple finite modulation wavevectors $\mathbf{Q}_i$ are identified, as exemplified in the main text.

\begin{figure}[htbp]
    \centering
    \includegraphics[width=\linewidth]{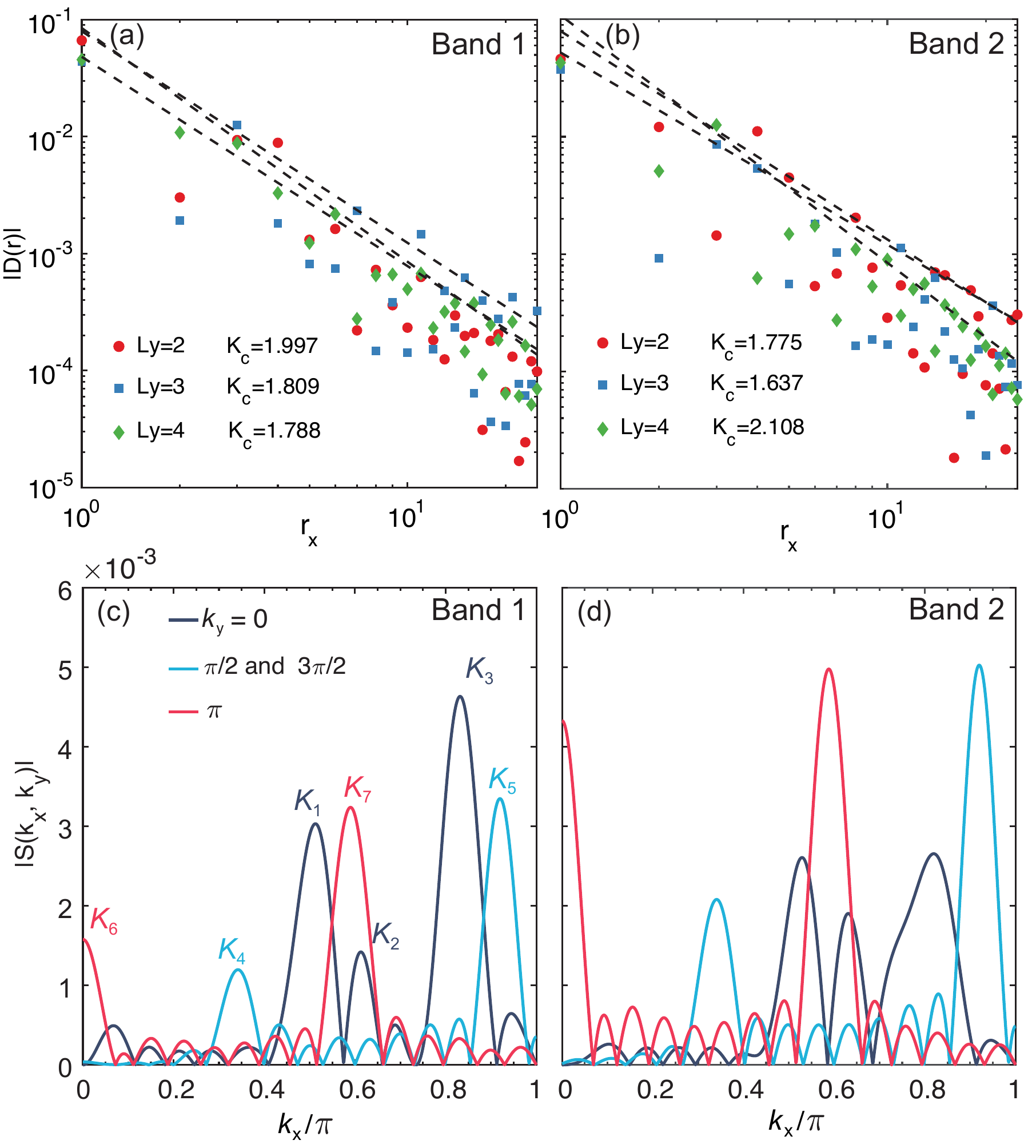}
    \caption{(a-b) Intraband density correlation $|D(r)|$ for $L_y = 2,3,4$ with power-law fits $|D(r)| \sim r_x^{-K_c}$. The dashed lines are fitting lines. (c-d)
    Density structure factor $|S(\mathbf{k})|$ in the reduced Brillouin zone 
             ($k_x \in [0,\pi]$) exploiting $|S(k_x,k_y)| = |S(-k_x,k_y)|$. 
             Data at $k_y = \pi/2$ and $k_y = 3\pi/2$ (blue curves) are scaled 10$\times$ for visual clarity. To minimize boundary effects, we exclude edge data in the Fourier analysis. Primary CDW peaks are labeled $K_1$ to $K_{7}$.}
    \label{Sfig: Density structure factor}
\end{figure}

We further examine the charge density correlations $D_{aa}(\mathbf{r})$ in Figure \ref{Sfig: Density structure factor}(a-b). These correlations also decay as a power law, but their exponents do not decrease significantly with increasing $L_y$. This contrasting behavior underscores the dominance of superconducting fluctuations over charge density wave fluctuations as the system approaches the 2D limit. The structure factor of the CDW, shown in Figure \ref{Sfig: Density structure factor}(c-d), allows us to identify the primary charge ordering wavevectors $\mathbf{K}_i$. Crucially, all major $\mathbf{K}_i$ can be expressed as differences of the PDW wavevectors, i.e., $\mathbf{K}_i = \mathbf{Q}_j - \mathbf{Q}_k$, within a numerical tolerance of $0.02\pi$. This exact correspondence provides strong evidence that the observed CDW is not an independent order but a secondary order parameter generated by the primary PDW order.

